\documentclass[aps,prc,eprint,showpacs,preprintnumbers, amsmath,amssymb,floatfix]{revtex4-2}

\usepackage{blkarray}

\usepackage{amsmath}
\usepackage{amssymb}
\usepackage{subfigure}
\usepackage{hyperref}
\usepackage{mathrsfs}
\usepackage{amsxtra}
\usepackage{mathrsfs}
\usepackage{amsfonts}
\usepackage{dsfont}
\usepackage{dcolumn}
\usepackage{bm}
\usepackage{graphicx}
\usepackage{epstopdf}
\usepackage{txfonts}
\usepackage{braket}
\usepackage{comment}

\usepackage{tikz} 
\usetikzlibrary{shapes.geometric, arrows} 
\tikzstyle{arrow} = [thick,->,>=stealth] 
\tikzset{mynode/.style={inner sep=2pt,fill,outer sep=0,circle}}

\begin{document}

\title{Towards a unified treatment of $\Delta S=0$ parity violation in low-energy nuclear processes} 

\author{Susan~Gardner}
\email{gardner@pa.uky.edu}
\affiliation{ Department of Physics and Astronomy, University of Kentucky, Lexington, KY 40506-0055, USA}

\author{Girish Muralidhara}
\email{girish.muralidhara@uky.edu}
\affiliation{ Department of Physics and Astronomy, University of Kentucky, Lexington, KY 40506-0055, USA}

\begin{abstract}
We revisit the unified treatment of low-energy hadronic parity violation espoused by 
Desplanques, Donoghue, and Holstein 
to the end of an 
{\it ab initio} treatment of parity violation in low-energy nuclear processes
within the Standard Model. We use our improved effective Hamiltonian 
and precise non-perturbative assessments of the quark charges of the nucleon 
within lattice QCD to make new assessments of the parity-violating meson-nucleon
coupling constants. Comparing 
with recent, precise 
measurements of
hadronic parity violation in few-body nuclear reactions, we find
improved agreement with these experimental results, though some 
tensions remain.
We thus note 
the broader problem of comparing
low-energy constants from nuclear and few-nucleon systems, 
considering, 
too, unresolved theoretical issues in 
connecting an {\it ab initio}, effective Hamiltonian 
approach to 
chiral effective theories. We note 
how future experiments 
and lattice QCD studies 
could sharpen 
the emerging picture, promoting the study of hadronic parity 
violation as a laboratory for testing ``end-to-end'' theoretical descriptions of 
weak processes in hadrons and nuclei at low energies. 
\end{abstract}

\maketitle

\section{Introduction}
\label{sec:Introduction}

In spite of decades of research, hadronic parity violation in flavor non-changing processes 
remains poorly understood~\cite{Desplanques:1979hn,Adelberger:1985ik,Haxton:2013aca,Schindler:2013yua,Gardner2017paradigm,deVries:2020iea}. The pertinent body of experimental work involves the low-energy 
interactions of hadrons and nuclei, so that we are compelled to 
address the interplay
of the physics of the weak interaction and of 
nonperturbative strong dynamics. 
Ultimately 
we hope that this problem can be largely conquered once 
the direct computation of two-nucleon matrix elements of a suitable effective Hamiltonian within 
lattice QCD (LQCD) becomes possible~\cite{Nicholson:2021zwi}, 
though, as we shall see, there are further issues to address. 
As an 
interim step, we revisit the unified treatment 
of hadronic parity violation by 
Desplanques, Donoghue, and Holstein (DDH)~\cite{Desplanques:1979hn}. 
There, the description of low-energy hadronic parity violation is 
framed within an one-meson-exchange model, and DDH show that it is possible to 
compute the appropriate meson-nucleon coupling constants
starting from the Standard Model (SM) Lagrangian. 
Since that early work, powerful field theoretic 
treatments exploiting the low-energy symmetries of QCD have
been developed and applied to the analysis of hadronic parity violation~\cite{Zhu:2004vw,Phillips:2008hn,Schindler:2009wd,deVries:2013fxa,Schindler:2013yua,deVries:2014vqa,deVries:2015pza,deVries:2020iea}. 
Yet in these
chiral effective field theory treatments, organized in terms of hadronic degrees of
freedom, 
the effective couplings are determined from experiment, and the 
underlying theoretical connection to QCD and the SM 
is lost. 
We note, however, nascent work that 
would compute the parity-violating pion-nucleon constant in an {\it ab initio} way~\cite{Wasem:2011tp,Feng:2017iqb,Sen:2021dcb}. 
Here we assess the current status of this problem by revisiting and updating
the treatment of DDH. Namely, we employ our improved effective Hamiltonian~\cite{Gardner:2022mxf}
to compute the parity-violating meson-nucleon
coupling constants, using the factorization approximation (as it
is now employed~\cite{Bauer:1986bm})
and LQCD assessments of the quark flavor charges of the 
nucleon~\cite{Aoki:2021kgd}. 
Our particular purpose is to see how these updated assessments combine
to confront the constraints on these parameters from precise 
experimental measurements of 
hadronic parity violation in few-body nuclear systems, 
namely, from the NPDGamma \cite{NPDGamma:2018vhh} 
and n3He \cite{n3He:2020zwd} collaborations 
that measure the parity-violating asymmetry from neutron-spin
reversal 
in the 
$\vec n + p \to d  + \gamma$ and in $\vec n\, + ^3{\rm He} \to t\, + p$ reactions, respectively. 

The 
NPDGamma 
measurement 
is particularly sensitive to the parity-violating 
pion-nucleon coupling, whereas that made by 
the n3He collaboration 
also probes four-nucleon contact interactions of isoscalar and
isovector character, 
which we interpret in terms of contributions from 
vector-meson exchanges between nucleons. 
Much of the past theoretical effort has 
concentrated 
on studying 
charged pion-nucleon interactions, due to a 
longstanding notion of its dominance in 
hadronic-parity-violating observables~\cite{Desplanques:1979hn}. 
However, noting the non-observation of 
parity violation in $^{18}$F radiative decay~\cite{Haxton:1981sf,Adelberger:1983zz,Page:1987ak}, 
and thus finding no clear 
sign of this dominance, and 
with the 
direct theoretical analysis of nucleon-nucleon (NN) amplitudes 
in pionless effective field theory (EFT)
in the large number of colors ($N_c$) limit 
showing 
that isoscalar and isotensor interactions 
should play driving 
phenomenological 
roles~\cite{Phillips:2014kna,Schindler:2015nga,Gardner2017paradigm,NPDGamma:2018vhh,n3He:2020zwd}, we believe the 
contributions from all isosectors should be 
computed. 
Earlier studies of QCD evolution effects 
have either made calculational approximations~\cite{Desplanques:1979hn,Dubovik:1986pj,Tiburzi:2012hx},
or focused on the isovector case~\cite{Dai:1991bx,kaplan1993analysis,Tiburzi:2012xx}.
We note, for example, that the original estimates of parity-violating 
meson-nucleon couplings were performed with a low-energy Hamiltonian 
built using phenomenological 
$K$-factors to account for 
QCD evolution effects on weak processes~\cite{Desplanques:1979hn}. 
In this work, we employ 
our low-energy effective Hamiltonian~\cite{Gardner:2022mxf}, which makes a complete
renormalization group evolution in leading-order QCD,  
with matching across heavy-flavor thresholds, 
to give a unified treatment of all three isosectors in order to compare 
their contributions to recent experimental measurements. 

Powerful searches for physics beyond the SM can be made through 
low-energy, precision measurements of symmetry-breaking 
effects in nucleons and nuclei~\cite{Cirigliano:2013lpa,Brambilla:2014jmp}. 
For example, in the case of searches for
permanent electric dipole 
moments, 
for neutrinoless
double $\beta$ decay, or 
for $\mu \to e$ conversion
on nuclear targets, the expected SM contribution
is either negligibly small with respect to current
experimental sensitivities 
or altogether absent. 
Thus the discovery of significantly 
non-zero results in these systems 
would signal the existence
of physics beyond the
SM. Here theory is key 
to assessing the relative sensitivity of 
different nuclear systems to the effects of interest. 
Theory is also essential to 
the interpretation of a non-zero 
experimental result or limit in terms of the 
parameters of an underlying 
new physics model --- and, more broadly, to using 
the experimental limit to estimate a lower bound
on the energy scale of new physics, assuming that 
it lies beyond the weak scale. 
A theoretical 
analysis that connects the scale of new physics to that
of the pertinent low-energy experiments 
requires the consideration of multiple 
physical scales, and 
``end-to-end'' effective-field-theory treatments
are being developed 
to accomplish that~\cite{Cirigliano:2018yza,Haxton:2022piv,Cirigliano:2022rmf}. 
In this context, we 
believe QCD studies of 
hadronic parity violation have a crucial role to play in the 
benchmarking of these treatments, because its observables
are not only nonzero within the SM but also, given the success
of the SM in describing ultra-low energy, parity-violating 
electron-nucleon interactions~\cite{Carlini:2019ksi}, 
new-physics effects presumably play a subdominant role. 
Thus the comparison of theory and experiment in hadronic parity violation 
provides a welcome test of the overall theoretical framework, as 
such tests possess aspects common to 
new-physics searches as well.

In this paper, we embark on this program by 
determining the parity-violating meson-nucleon 
coupling constants at a renormalization scale of 2 GeV, and, 
as we shall detail, we
find improved agreement with the experimental results. 
Although better agreement speaks to 
progress, our longer-term goals are 
to refine our results to higher precision and also 
to evolve our description to still smaller scales. 
We note that the parity-violating meson-nucleon couplings, 
and, generally, the low-energy constants associated
with the operators of an effective field theory are 
not in themselves observables and can be expected
to depend on the renormalization scale. In 
this paper we discuss various assessments of the 
parity-violating pion-nucleon coupling constant 
from this perspective, as it is the most precisely 
determined. 
%
Generally, we anticipate 
difficulties can arise both from the gap 
between the lowest scale to which we can potentially
apply perturbative QCD accurately and the highest scale to which 
we can employ chiral
perturbation theory, 
as well as from the effects 
of the massive charm quark. 
The latter affects the splay of 
operators that can appear,  
even if the charm quark
is still active~\cite{Buchalla:1995vs}, as we have 
developed explicitly in the $\Delta S=0$ case~\cite{Gardner:2022mxf}. 
Moreover, the truncation error from matching 
a four to three-flavor theory, at a fixed order of perturbation theory, at
charm threshold can be significant, as studied in $K\to\pi\pi$ 
decay~\cite{Tomii:2020smd,Tomii:2022vxb}, where we refer to Ref.~\cite{Blum:2022wsz}
for a broader discussion. 
In this paper we comment 
on how some of these effects can impact our results.

We conclude this section with an outline of the rest of the paper. 
In Sec.\ref{sec:II}, we recapitulate the outcomes of our 
effective weak Hamiltonian computation~\cite{Gardner:2022mxf} that are pertinent here. 
In Sec.\ref{sec:III} we discuss 
the factorization approximation and its validity. In Sec.\ref{sec:IV} we employ these results to compute the parity-violating meson-NN coupling constants. 
In Sec. \ref{sec:V} we 
compare our results with the parameters extracted from experiments
and discuss the perspectives they offer, 
and we offer a concluding summary and outlook in Sec. \ref{sec:VI}.

\section{Effective Hamiltonian and Wilson Coefficients}
\label{sec:II}

The effective Hamiltonian for hadronic parity violation 
at a particular energy scale is defined in terms of four-quark operators and corresponding Wilson coefficients. 
In evolving the theory from one energy scale to another, such as from the $W$ mass to scale $\mu$, 
the Wilson coefficients follow the relation: 
\begin{equation}\label{WC RG}
        \Vec{C}(\mu) = \exp\left[\int_{g_s(M_W)}^{g_s(\mu)}dg \frac{\gamma^T(\mu)}{\beta(g_s)}\right]\Vec{C}(M_W) \,\quad \rm{with} \quad \beta(g_s) = -\frac{g_s^3}{48\pi^2}(33-2n_f), 
\end{equation}
where we work in leading-order (LO) QCD, noting that 
the anomalous dimension matrix $\gamma$ 
arises from the LO
QCD mixing of the operators. We 
refer to Ref.\cite{Gardner:2022mxf} for all details. 
Allowing the effective theory to flow from the $W$ mass scale to 
hadronic energy scales, with $\mu = 2\, \rm GeV$, results in the Hamiltonian:
\begin{equation}
     \mathcal{H}_{\rm eff}^{\rm PV}(2 \,\rm GeV) = \frac{G_Fs_w^2}{3\sqrt{2}}\sum_{i=1}^{12} C_i(2 \,\rm GeV)\, \Theta_i \,,
\end{equation}
where $\Theta_i$ are four-quark operators. Twelve such operators form a closed set under LO QCD mixing and thus describe 
the theory of hadronic parity nonconservation, 
for all three isosectors, where we refer to App.~\ref{full ops} for a complete list.   
Moreover, we use the weak-mixing angle $\theta_W$ with 
$\sin^2\theta_W\equiv s_w^2=0.231$ and the Fermi constant
$G_F=1.166 \times 10^{-5}\, \rm{GeV^{-2}}$~\cite{Zyla:2020zbs}. Just below the W mass scale, the effects of QCD are negligible,
and we can collect the Wilson coefficients by summing the tree-level W and $Z^0$ exchange contributions in $\Delta S = 0$ quark interactions, giving 
$\Vec{C}(M_W) = (1, 0, 0, 0, -3.49, 0, 0, 0, -13.0\cos^2\!\theta_c, 0, -13.0\sin^2\!\theta_c, 0)$, with the Cabibbo angle given by $\sin \theta_c=0.2253$. Upon performing the RG flow to 2 GeV using Eq.\ref{WC RG}, we have~\cite{Gardner:2022mxf}
\begin{equation} 
        \!\!\!\!\!\!\!\!\!\!\Vec{C}(2 \,\rm GeV)\! = \!\begin{pmatrix} 
            1.09 &[1.17 \dots 1.06] [1.08 \dots 1.04] &[1.07] [1.06]\\
            0.018 &[0.014 \dots 0.021] [0.033 \dots 0.006] &[-0.006] [-0.006]\\
            0.199 &[0.321\dots 0.133] [0.193 \dots 0.127] &[0.158] [0.153]\\
            -0.583 &[-0.990 \dots -0.385] [-0.571 \dots -0.374] &[-0.460] [-0.456]\\
            -4.36  &[-4.99 \dots -4.05] [-4.34 \dots -4.03] &[-4.16] [-4.14]\\
            1.72 &[2.63 \dots 1.19] [1.67 \dots 1.16]  &[1.40] [1.36]\\
            -0.170 &[-0.288 \dots -0.110] [-0.165 \dots -0.105] &[-0.134] [-0.129]\\
            0.332 &[0.496 \dots 0.235] [0.322 \dots 0.225] &[0.275] [0.268]\\
            -16.2 &[-18.6 \dots -15.0] [-16.1 \dots -15.0] &[-15.48] [-15.4] \\
            6.38 &[9.76 \dots 4.44] [6.22 \dots 4.30]  &[5.19] [5.05] \\
            -16.2 &[-18.6 \dots -15.0] [-16.1 \dots -15.0] &[-15.48] [-15.4] \\
            6.38  &[9.76 \dots 4.44] [6.22 \dots 4.30] &[5.19] [5.05]
        \end{pmatrix} \,,
        \label{eq:fullheff}
\end{equation}
where the last four entries should be multiplied by factors of 
 $\cos^2\!\theta_c,  \cos^2\!\theta_c,  \sin^2\!\theta_c$, and $\sin^2\!\theta_c$, respectively. 
The primary result
is given by the leftmost column of numbers. The other columns illustrate the uncertainties in the computation. 
In the central column, the left set shows the ranges of Wilson coefficients that result 
in the $N_f=2+1$ theory for renormalization scales of $\mu=1-4 \,{\rm GeV}$ and the right set 
shows them in the $N_f=2+1 +1$ theory with $\mu=2-4 \,{\rm GeV}$. The rightmost column gives 
Wilson coefficients if the $\alpha_s$ running and matching is computed at NLO (left) and NNLO (right). 

For the present work, it is useful to make
the different isosector contributions explicit and separated as
\begin{equation}
    \mathcal{H}_{\rm eff}^{\rm PV}(2 \,\rm GeV) = \mathcal{H}_{\rm eff}^{\rm I=1}(2 \,\rm GeV) + \mathcal{H}_{\rm eff}^{\rm I=0\oplus2}(2 \,\rm GeV).
\end{equation}
Isovector ($I=1$) Wilson coefficients at high and low energies are: 
$\Vec{C}^{I=1}(M_W) = (1, 0, 0, 0, 3.49, 0, 3.49, 0, -13.0\cos^2\!\theta_c, 0)$ and
\begin{equation}\label{vec wil coef}
        \!\!\!\!\!\!\!\!\!\!\Vec{C}^{I=1}(2 \,\rm GeV)\! = \!\begin{pmatrix} 
            1.09 &[1.17 \dots 1.06] [1.08 \dots 1.04] &[1.07] [1.06]\\
            0.018 &[0.014 \dots 0.021] [0.033 \dots 0.006] &[-0.006] [-0.006]\\
            0.199 &[0.321\dots 0.133] [0.193 \dots 0.127] &[0.158] [0.153]\\
            -0.583 &[-0.990 \dots -0.385] [-0.571 \dots -0.374] &[-0.460] [-0.456]\\
            4.36  &[4.99 \dots 4.05] [4.34 \dots 4.03] &[4.16] [4.14]\\
            -1.72 &[-2.63 \dots -1.19] [-1.67 \dots -1.16]  &[-1.40] [-1.36]\\
            4.36  &[4.99 \dots 4.05] [4.34 \dots 4.03] &[4.16] [4.14]\\
            -1.72 &[-2.63 \dots -1.19] [-1.67 \dots -1.16]  &[-1.40] [-1.36]\\
            -16.2 &[-18.6 \dots -15.0] [-16.1 \dots -15.0] &[-15.48] [-15.4] \\
            6.38 &[9.76 \dots 4.44] [6.22 \dots 4.30]  &[5.19] [5.05] \\
        \end{pmatrix} \,,
\end{equation}
where the last two entries should be multiplied by a factor $\sin^2\!\theta_c$ and 
the error estimates are defined as in Eq.~(\ref{eq:fullheff}). Wilson coefficients for the $I=0 \oplus 2$ sector 
at high and low energies are: 
$\Vec{C}^{I=0\oplus2}(M_W) = (-1, 0, 0, 0, -3.49, 0, 0, 0, -13.0\cos^2\!\theta_c, 0)$ and  
\begin{equation} 
        \!\!\!\!\!\!\!\!\!\!\Vec{C}^{ I=0\oplus2}(2\,\rm GeV)\! = \!\begin{pmatrix} 
            -1.09 &[-1.17 \dots -1.06] [-1.08 \dots -1.04] &[-1.07] [-1.06]\\
            -0.018 &[-0.014 \dots -0.021] [-0.033 \dots -0.006] &[0.006] [0.006]\\
            -0.199 &[-0.321\dots -0.133] [-0.193 \dots -0.127] &[-0.158] [-0.153]\\
            0.583 &[0.990 \dots 0.385] [0.571 \dots 0.374] &[0.460] [0.456]\\
            -4.36  &[-4.99 \dots -4.05] [-4.34 \dots -4.03] &[-4.16] [-4.14]\\
            1.72 &[2.63 \dots 1.19] [1.67 \dots 1.16]  &[1.40] [1.36]\\
            -0.170 &[-0.288 \dots -0.110] [-0.165 \dots -0.105] &[-0.134] [-0.129]\\
            0.332 &[0.496 \dots 0.235] [0.322 \dots 0.225] &[0.275] [0.268]\\
            -16.2 &[-18.6 \dots -15.0] [-16.1 \dots -15.0] &[-15.48] [-15.4] \\
            6.38 &[9.76 \dots 4.44] [6.22 \dots 4.30]  &[5.19] [5.05] \\ \,,
        \end{pmatrix}
        \label{sca ten wil coef}
\end{equation}
where the last two entries should be multiplied by a factor $\cos^2\!\theta_c$ and the error estimates are defined as in Eq.~(\ref{eq:fullheff}).
Although Eqs.~(\ref{eq:fullheff}, 
\ref{vec wil coef}, \ref{sca ten wil coef}) 
appeared in our earlier paper~\cite{Gardner:2022mxf}, 
we have included them here
to make our presentation self-contained.

\section{Factorization Approximation}
\label{sec:III}
The effective Hamiltonian presented in the previous section can be used in the computation of various parity-violating
meson-nucleon coupling constants of isospin $I$: $h_M^I$. These parameters are 
introduced to quantify the observable effects of hadronic parity violation via the phenomenological Hamiltonian 
of Ref.~\cite{Desplanques:1979hn}, $\mathcal{H}_{\rm DDH}$. By matching the quark-level and hadron-level matrix elements via 
$\langle M N' | \mathcal{H}_{\rm eff}^I  | N\rangle = \langle M N' | \mathcal{H}_{\rm DDH}  | N\rangle$, these couplings can be estimated. The main challenge in this is determining the
quark-level matrix elements 
$\langle M N' | \mathcal{H}_{\rm eff}^I  | N\rangle$ 
involving hadrons. This task is significantly simplified within the factorization, or vacuum
saturation, approximation, in which the hadronic
matrix element of the four-quark operator is computed
as the product of the hadron matrix elements of 
each current. The factorization approximation is heuristic, 
though its use can be justified {\it a posteriori}
with experimental data, if not {\it a priori} on theoretical grounds, except in special cases. 
The difference in the matrix-element computation 
of the full 4-quark operator and that of its 2-quark pieces
is termed 
a non-factorizable contribution. This difference is 
not well-known in general, 
and its outcome depends on the matrix element chosen.

To our knowledge, the factorization 
approximation was first studied 
in the context of hadronic parity 
violation~\cite{Michel:1964zz,Fischbach:1969gyc}; 
in particular, 
the matrix element of a parity-violating four-quark operator 
to yield a neutral vector meson from a nucleon state is 
thus written in the form 
\begin{equation}
     \bra{V N'}(\bar{q_1}q_2)_V(\bar{q_3}q_4)_A\ket{N} = \bra{V}(\bar{q_1}q_2)_V\ket{0}\bra{N'}(\bar{q_3}q_4)_A\ket{N} \,.
\end{equation}
Factorization has also been broadly employed in analyses of 
hadronic weak decays, with the first application being to the
computation of so-called tree graphs, arising 
from partially disconnected
intermediate states, and their contribution to the 
$|\Delta I|=1/2$ rule in $K\to 2\pi$ and 
$K\to 3\pi$ decay~\cite{Feynman:1964FA,Haan:1970yi}. 
With further developments, the factorization approximation 
has been used to yield predictions for the 
exclusive decays of charmed mesons~\cite{Fakirov:1977ta,Cabibbo:1977zv}, 
compared against experimental 
data~\cite{Bauer:1984zv,Bauer:1986bm}, and 
applied to the B-meson system, 
in which extensive tests become possible through
the rich selection of possible hadronic 
final states~\cite{Ali:1998eb,Cheng_Tseng1998PhRvD..58i4005C,Diehl_Hiller2001JHEP...06..067D}. 
Under certain conditions, factorization has been shown 
to work extremely well. To that end 
we consider the specific example of 
B-meson decays to heavy-light final states, for 
which factorization has been shown to exist in QCD 
in leading inverse power in the heavy quark 
mass~\cite{Beneke:2000ry,Bauer:2001cu} 
--- assuming that both $b$ and $c$ quarks are heavy. 
Tests of these predictions, and of factorization more
generally, come from the study of 
$B\to D^{(*)}_{(s)} (\pi,K)$ decays~\cite{Fleischer:2010ca},
particularly 
the comparison of the theoretical decay rates
with experiment, yielding excellent agreement. 
These decays include both vector and pseudoscalar final 
states and probe the 
color-suppressed (C) and exchange (E) topologies, 
in addition to the color-allowed tree (T) contribution. 
%
For example, 
a test derived from ${\bar B}_d^0 \to D^+ \rho^-$ and 
${\bar B}_d^0 \to D^+ \pi^-$ branching ratio data, which 
is sensitive to both the T and E topologies, probes factorization
to a precision of 10\%, and the authors note
that they could not resolve any nonfactorizable 
effects within the current experimental precision, 
which could be as small as 5\% in some cases~\cite{Fleischer:2010ca}. 
%
%
In contrast, 
in meson decays to light final states, 
the energy release is generally much 
larger, admitting the possibility of rescattering with intermediate-state 
hadronic resonances and thus yielding contributions beyond the factorization 
approach. 
Empirical uncertainties
in $B\to \pi\pi,\pi K$ decays are still large enough to 
preclude such precise tests~\cite{Fleischer:2018bld}. In this class of
decays, an outstanding problem has been that of understanding the
pattern of amplitudes in 
$K\to\pi\pi$ decay, for which a marked dominance of the $I=0$ final 
state amplitude over the $I=2$ amplitude is observed, with 
roughly only a factor of 2 of the empirical ratio 
${\rm Re} A_0/ {\rm Re} A_2 \simeq 22.5$ in the isospin 
limit~\cite{Gardner:2000sb} 
coming from the 
perturbative Wilson coefficients and a simple factorization of the 
hadronic matrix elements. Although the problem has long been 
attributed to an unidentified enhancement of the $I=0$ 
amplitude~\cite{Buchalla:2001ux,Cirigliano:2011ny}, to 
which a role for the $\sigma(500)$ resonance has been 
argued~\cite{Shabalin:1987jf,Morozumi:1990ic}, 
LQCD studies have now shown that a numerical resolution 
of the $|\Delta I|=1/2$ puzzle~\cite{RBC:2020kdj} 
includes a significant cancellation 
of two tree-level operators that contribute to the $I=2$ amplitude in 
$K\to\pi\pi$ decay~\cite{boyle2013emerging,Blum:2015ywa,RBC:2020kdj}. In the
factorization treatment the two contributions have the same sign, 
showing it to be inconsistent. We note that an opposite
relative sign also emerged in earlier non-lattice work using 
chiral 
perturbation theory and a large $N_c$ analysis~\cite{Bardeen:1986vz,Pich:1995qp}. 
The analysis of $K\to \pi\pi$ decays reveals features that 
do not occur in our analysis of the parity-violating meson-nucleon couplings. 
In particular, 
since QCD dynamics are flavor-blind, we believe
that the existing 
factorization tests in heavy to heavy-light transitions
do have bearing on our $N\to N M$ analysis, supporting
our results because the 
kinematics of the process does not support 
the existence of factorization-violating
resonances. 
We would like to 
emphasize that we employ
the factorization approximation specifically for 
the computation of the parity-violating meson-nucleon coupling constants. 
The issue of non-perturbative effects beyond the DDH model, which could be studied within the framework of 2N matrix elements within LQCD remains. 
Moreover, the theoretical improvements we have made are specific to the computation of the meson-nucleon coupling constants. 
To put this in context we now turn to the analysis of 
DDH~\cite{Desplanques:1979hn}.

The early landmark study of hadronic parity violation 
by DDH~\cite{Desplanques:1979hn} is critical
of the use of factorization, 
and that assessment
has appeared to hold sway despite later work 
suggesting that non-factorizable effects are 
subdominant~\cite{Dubovik:1986pj}. 
In regards to the comparative study of DDH~\cite{Desplanques:1979hn}, 
both the factorization approximation computations 
and the quark model estimates to which they were compared 
employed uncontrolled approximations, and poorly 
known inputs, so that inferring a deficiency in the factorization
approximation itself from differences in such predictions 
is not a reliable conclusion. Moreover, what 
DDH term ``factorization'' is not the same procedure as
has been employed in the literature since the
late 1980's~\cite{Bauer:1986bm}. In their Fig.~1
they present three different quark flow topologies for the 
parity-violating meson-nucleon couplings and 
note that 
``factorization'' is associated with the production of 
a color-singlet meson emerging
as the result of $Z^0$ exchange at tree level exclusively. 
Moreover, different paths to computing 
factorized hadronic matrix
elements are employed~\cite{Desplanques:1979hn}. 
We, rather, have followed 
the now standard practice of applying 
a Fierz transformation to a four-quark operator 
to expose the quark currents with the flavor content needed
to realize a particular hadronic final state, so that 
we factorize the matrix elements of the four-quark operators 
into products of the matrix elements of the associated
quark-level currents.
In so doing the matrix elements of 
our LO weak Hamiltonian can generate all the 
pictorial contributions in Fig.~1 of ~\cite{Desplanques:1979hn}, depending
on the meson to be produced. For example, their 
Fig. 1b can follow from multi-quark (in excess of
three) Fock states of 
the nucleon, as associated with the strange quark
axial charge of the nucleon, which is pertinent
to the assessment of the vector-meson-nucleon
couplings. 

In making our assessments, we have employed the 
%
recent, precision LQCD 
computations of the quark-flavor (scalar, axial) charges of the nucleon~\cite{Aoki:2021kgd}, and we regard that as a great improvement
over 
the poorly controlled 
flavor-symmetry-based estimates
used 
throughout the literature 
in the past, 
as we discuss in Sec.~\ref{sec:IV}. This is key to a sharpened picture 
of the role of strange quarks, which have been a source of great
uncertainty~\cite{Dai:1991bx,Meissner:1998pu}. 
Thus 
greatly improved assessments of the factorized matrix elements 
in the nucleon sector are now possible. 
Turning to the parity-violating 
$n\to p (\pi^\pm, \rho^\pm, \omega)$ transition matrix elements, we note
these processes, though now mediated by $Z^0$ exchange, also
contain quark flow topologies of the same forms studied 
by Ref.~\cite{Fleischer:2010ca}, and the kinematics 
of these transition matrix elements is also 
compatible 
with that of 
the heavy-quark/hadron limit they employ. 
Thus we regard those 
tests of factorization in hadronic B meson decays,
which speak to its success in that context, 
as also acting in support of our own analysis. 
In the next section we 
use the factorization approximation with input from state-of-the-art lattice QCD results to determine
the parity-violating meson-nucleon coupling constants. 

%

\section{Estimates of the parity-violating meson-nucleon coupling constants}
\label{sec:IV}

In this section, firstly we flesh out 
the calculation of $h_{\pi}^1$ in Ref.\cite{Gardner:2022mxf}, particularly emphasizing and discussing the different 
input choices made in arriving at this result. Then, 
we turn to estimating 
the remaining meson-nucleon couplings. Phenomenologically, the pion contribution to hadronic parity violation
with coupling $h_{\pi}^1$ is 
\begin{equation}
    \mathcal{H}^{\pi}_{\rm DDH} = i h^1_{\pi}(\pi^{+}\bar{p}n-\pi^{-}\bar{n}p) \,.
\end{equation}
Matching the quark and hadron-level matrix elements we have 
\begin{equation}
    -i h_{\pi}^1 \bar{u}_n u_p = \bra{n\pi^{+}}\mathcal{H}_{\rm eff}^{I=1}\ket{p}\,,
    \label{pimatch}
\end{equation}
where $u_N$ with $N\in p,n$ is a Dirac spinor. Employing
the Fierz identities, where we note the useful compilation of Ref.~\cite{Nieves:2003in}, $\Theta_i^{I=1}$ operators within the Hamiltonian are rearranged to yield
scalar-pseudoscalar contributions. Using the definition $\langle 0 | (\bar d u)_A (0) |  \pi^+ (p) \rangle = i p^\mu f_\pi$ 
and the result
\begin{equation}
    \bra{\pi^+} (\bar{u}   \gamma_{5} d) \ket{0} = \frac{m_{\pi}^2 f_{\pi}}{i(m_u+m_d)} \,,
\end{equation}
we obtain the equation connecting the pion-nucleon coupling to the Wilson coefficients in ${\cal H}_{\rm eff}^{I=1}$: 
\begin{equation}
     h_{\pi}^1 \bar{u}_n u_p=\frac{2G_Fs_w^2}{3\sqrt{2}}\left(\frac{C_{1}^{I=1}}{3}+C_{2}^{I=1}-\frac{C_{3}^{I=1}}{3}-C_{4}^{I=1}\right)\frac{m_{\pi}^2 f_{\pi}}{(m_u+m_d)}
     \bra{n}\bar{d}u\ket{p}\,.
     \label{evalhpi1}
\end{equation}
In its numerical evaluation, we use the 
isovector scalar charge $g_s^{u-d}$ computed within lattice QCD (LQCD)~\cite{Aoki:2021kgd}, where 
$\bra{n}\bar{d}u\ket{p}\equiv g_s^{u-d}\bar{u}_nu_p$.
Modern LQCD calculations are 
``unquenched'' 
so that the effects 
of the light sea quarks are allowed to appear, noting that these are characterized by $N_f$, the number of dynamical
quark flavors in the simulation. As per Ref.~\cite{Aoki:2021kgd}, we suppose simulations with $N_f=2+1+1$ are 
more 
realistic but that $N_f=2+1$ simulations 
are typically more precise. 
The evaluation of Eq.~(\ref{evalhpi1}) is sensitive to the precise value 
of $m_\pi^2/(m_u + m_d)$, where 
the light quark masses are evaluated
in LQCD. This ratio gives a large enhancement, and its assessment should be made with care. 
Here $m_\pi=135\,\rm MeV$, because the LQCD simulations used do not include electromagnetism, 
and 
the charged-pion 
decay constant $f_\pi=130\,\rm MeV$. As for the light quark masses, 
it is appropriate to use the renormalization-group-invariant (RGI) mass
$(m_u+m_d) = 2(4.695 (56)_m (54)_\Lambda\,\rm MeV)$ for $N_f=2+1$ ~\cite{Aoki:2021kgd}, an appealing choice because it is scale and scheme independent, thus
avoiding 
extreme sensitivity to the 
choice of scale. In this case, combining errors in quadrature implies 
$m_\pi^2/(m_u + m_d) = 1941 (32)\,\rm MeV$, 
whereas using the result from a $N_f=2+1$ simulation in 
the ${\overline{\rm{MS}}}$ scheme at a scale of $2\,\rm GeV$,  
$(m_u+m_d) = 2(3.381(40)\,\rm MeV)$~\cite{Aoki:2021kgd}
we find $2695 (32) \,\rm MeV$ for this ratio. 
(We note in this 
scheme at this scale that the PDG compilation recommends 
$(m_u+m_d) = 2(3.45^{+0.55}_{-0.15}\,\rm MeV)$~\cite{Zyla:2020zbs}; we note, too, 
$(m_u+m_d) = 2(3.75 (0.45)\,\rm MeV)$ using scalar sum rules
and chiral perturbation theory~\cite{Jamin:2006tj}.)
We can also assess it 
through the
use of the Gell-Mann--Oakes--Renner (GOR)
relation~\cite{Gell-Mann:1968hlm,Gasser:1983yg,Gasser:1984gg}. The GOR relation captures the pion mass with a correction of within 
a few percent~\cite{Jamin:2002ev,Bernard:2006gx,Bordes:2010wy,McNeile:2012xh}, where 
the concomitant quark condensate $B\equiv |\Sigma|/ F^2$, with 
$\Sigma = |\langle 0 | \bar u u | 0 \rangle|$ and $F$ the pion decay constant in the chiral limit, 
can all be computed in LQCD. Using Ref.~\cite{Aoki:2021kgd} to compute $B$ from 
$\Sigma$ and $F$, in the SU(2) chiral limit
and $N_f=2+1$ we have, assuming the errors are uncorrelated, 
$2560 (240)\,\rm MeV$, whereas in the SU(3) chiral limit we have 
$2280 (280) \,\rm MeV$, 
a difference reflecting the role of the strange sea quarks in
its numerical evaluation. 
The result with the RGI quark mass has been employed in what follows.
Turning to the isovector quark scalar charge of the nucleon, $N_f=2+1$
result: 
we use 
$g_s^{u-d}=1.06(10)(06)_{sys}$~\cite{park2021precision},
noting that this 
compares favorably with the result 
$g_s^{u-d}=1.02(11)$ determined from strong-isospin breaking in the nucleon mass from 
LQCD~\cite{Gonzalez-Alonso:2013ura}, whereas the SU(3) estimate in Ref.~\cite{kaplan1993analysis}
yields $0.6$. 
Finally,
\begin{equation}
 h_{\pi}^1  = (3.06 \pm 0.34  + \left(\stackrel{{+1.29}}{{}_{-0.64}}\right) + 0.42 + (1.00))\times 10^{-7} \,,
 \label{eq:hpi1}
\end{equation}
where the error estimates come, respectively, 
from the LQCD inputs employed, the change in the Wilson coefficients over (i) a scale variation 
of $1-4\,{\rm GeV}$ and (ii) higher-order corrections in $\alpha_s$ as per Eq.~(\ref{vec wil coef}), 
and, finally, the estimates of the accuracy of Eq.~(\ref{evalhpi1}) through the contribution 
to it from ${\cal O}(1/N_c)$ terms, which are noted in parentheses. 

We now turn to the assessment of other meson-nucleon coupling constants, starting with the 
remaining $I=1$ couplings. For the $\rho^0$ meson, e.g., $\bra{\rho^0 N}\mathcal{\cal H}_{\rm eff}^{I=1}\ket{N}=h_{\rho}^1\epsilon^{*\mu}_{\rho}(\bar{u}_N u_N)_A$.
With $ \bra{\rho^0}(\bar{u}u)_V-(\bar{d}d)_V\ket{0}\equiv \sqrt{2}\epsilon^{*\mu}_{\rho}f_{\rho}m_{\rho}$,
$m_\rho= 775.4 \, {\rm MeV}$~\cite{Zyla:2020zbs}, and $f_\rho=210 \,{\rm MeV}$~\cite{Ali:1998eb}
and using the quark axial charges of the nucleon from LQCD~\cite{Aoki:2021kgd}
\begin{align}\label{lqcd charges}
    \begin{split}
        &\bra{p}(\bar{u}u)_A\ket{p} = g_A^{u}(\bar{u}_p u_p)_A \,;\quad g_A^u = 0.777 (25)(30)
        \,\, [0.847 (18)(32)]\,,
        \\
        &\bra{p}(\bar{d}d)_A\ket{p} = g_A^{d}(\bar{u}_p u_p)_A \,;\quad g_A^d =-0.438 (18) (30) 
         \,\, [-0.407 (16)(18)] \,,
        \\
        &\bra{p}(\bar{s}s)_A\ket{p}= g_A^{s}(\bar{u}_p u_p)_A 
        \,;\quad g_A^s=-0.053 (8) \,\, [-0.035 (6)(7)] \,,
    \end{split}
\end{align}
in the $\overline{\rm MS}$ scheme at $\mu=2\,\rm GeV$ from  $N_f=2+1+1$~\cite{Lin:2018obj} [$N_f=2+1$~\cite{Liang:2021pql}] flavor 
simulations, we have 
\begin{equation}
   \begin{split}
   h_\rho^1
   =\frac{G_Fs_w^2}{3}f_\rho m_\rho \Bigg(\left(C_{3}^{I=1}+\frac{C_{4}^{I=1}}{3}\right)
   (g_A^u + g_A^d) 
    + \left(C_{3}^{I=1}+\frac{C_{4}^{I=1}}{3}+C_{7}^{I=1}+\frac{C_{8}^{I=1}}{3}+C_{9}^{I=1}+\frac{C_{10}^{I=1}}{3}\right) g_A^s\Bigg) \,,
  \end{split}
    \end{equation}
and with Eq.~(\ref{lqcd charges}) this yields 
\begin{equation}
         h^1_{\rho} = -0.294\pm 0.045  + \left(\stackrel{{0.014}}{{}_{-0.036}}\right) + 0.009 + (0.026))\times 10^{-7} \,,
\end{equation} 
For the $\omega$ meson, $\bra{\omega N}\mathcal{H}_{\rm eff}^{I=1}\ket{N}=h_{\omega\,N}^1\epsilon^{*\mu}_{\omega}(\bar{u}_Nu_N)_A$. 
With $\bra{\omega}(\bar{u}u)_V+(\bar{d}d)_V\ket{0}\equiv\sqrt{2}\epsilon^{*\mu}_{\omega}f_{\omega}m_{\omega}$, $m_\omega= 782.65 \,{\rm MeV}$~\cite{Zyla:2020zbs}, and $f_\omega=195 \,{\rm MeV}$~\cite{Ali:1998eb}, we have 
\begin{equation}
h_{\omega\,N}^1  
=\frac{G_Fs_w^2}{3}f_\omega m_\omega \Bigg(\left(C_{1}^{I=1}+\frac{C_{2}^{I=1}}{3}\right)\eta_N (g_A^u - g_A^d) +  \left(C_{9}^{I=1}+\frac{C_{10}^{I=1}}{3}\right)g_A^s \Bigg) \,,
\end{equation}
where $\eta=\pm 1$ for a proton or neutron state, respectively. 
With Eqs.(\ref{lqcd charges})
\begin{equation}
     h^1_{\omega\,p}= + 1.825\pm 0.111 + \left(\stackrel{{-0.047}}{{}_{0.125}}\right) -0.040 + (-0.020))\times 10^{-7}\,;\,
     h^1_{\omega\,n} = - 1.828\pm 0.112  + \left(\stackrel{{0.053}}{{}_{-0.134}}\right) + 0.043 + (0.000))\times 10^{-7} \,,
\end{equation}
where the difference in their magnitudes speaks to the role of charged-current effects. 
Similarly we can make use of $H_{\rm eff}^{I=0\oplus2}$ to determine 
$\bra{\omega N}\mathcal{H}^{I=0\oplus2}\ket{N}=h_{\omega}^0\epsilon^{*\mu}_{\omega}(\bar{u}_Nu_N)_A
$. Thus 
\begin{equation}
    \begin{split}
h_{\omega}^0 =\frac{G_Fs_w^2}{3} f_\omega m_\omega \Bigg(\left(C_{7}^{0+2}+\frac{C_{8}^{0+2}}{3}+C_{9}^{0+2}+\frac{C_{10}^{0+2}}{3}\right) (g_A^u + g_A^d)
     + \left(C_{1}^{0+2}+\frac{C_{2}^{0+2}}{3}+C_{7}^{0+2}+\frac{C_{8}^{0+2}}{3}\right) g_A^s\Bigg) \,,
    \end{split}
\end{equation}
and with Eqs.(\ref{lqcd charges}) this gives
\begin{equation}
     h^0_{\omega} = +0.270\pm 0.015  + \left(\stackrel{{-0.32}}{{}_{0.55}}\right) -0.202 + (1.148))\times 10^{-7} \,
\end{equation}
To determine the isocalar and isotensor $\rho$ couplings from ${\cal H}_{\rm eff}^{I=0\oplus2}$
we note from ${\cal H}_{\rm DDH}$~\cite{Desplanques:1979hn} that 
\begin{equation}\label{simul}
    h_{\rho}^{0}+\frac{1}{\sqrt{6}}h_{\rho}^{2} = h_{\rho^0}^{0\oplus 2}\,;\quad 
    \sqrt{2} h_{\rho}^{0} - \frac{1}{\sqrt{12}}h_{\rho}^{2} = h_{\rho^-}^{0\oplus 2} 
    \,.
\end{equation}
Computing $h^{0\oplus2}_{\rho^0}$, with $\bra{\rho^0 N}\mathcal{H}_{\rm eff}^{I=0\oplus2}\ket{N}=h_{\rho^0}^{0\oplus2}\eta_N \epsilon^{*\mu}_{\rho}(\bar{u}_Nu_N)_A$, 
\begin{equation}
      h_{\rho^0}^{0\oplus2} 
      =\frac{G_F s_w^2}{3}f_\rho m_\rho 
     \left(C_{5}^{I=0+2}+\frac{C_{6}^{I=0+2}}{3}-\frac{C_{9}^{I=0+2}}{6}-\frac{C_{10}^{I=0+2}}{2}\right)  (g_A^u - g_A^d)  \,,
\end{equation}
which, with Eqs.(\ref{lqcd charges}), implies
\begin{equation}
         h^{0\oplus2}_{\rho^0} = -7.55 \pm 0.46  + \left(\stackrel{{1.54}}{{}_{-2.76}}\right) + 1.00 + (-5.57))\times 10^{-7} \,.
\end{equation}
Computing $h^{0\oplus2}_{\rho^-}$, with  
$\bra{\rho^- p}\mathcal{H}_{\rm eff}^{I=0\bigoplus2}\ket{n}=h_{\rho^-}^{0\oplus2}\epsilon^{*\mu}_{\rho}(\bar{u}_Nu_N)_A$, 
noting $\bra{\rho^-}(\bar{d}u)_v\ket{0}=\epsilon^{*\mu}_{\rho}f_{\rho}m_{\rho}$, 
and using the quark isovector axial charge in LQCD in ${\overline {\rm MS}}$ at $2\,\rm GeV$ 
from a $N_f=2+1$~\cite{park2021precision} [$N_f=2+1+1$~\cite{Gupta:2018qil}] flavor simulation, namely, 
\begin{equation}
  \bra{p}(\bar{u}d)_A\ket{n} = g_A^{u-d}(\bar{u}_p u_n)_A; \quad g_A^{u-d} = 1.31(06)(05)_{\rm sys} \,[1.218 (25)(30)_{\rm sys}]
  \,,
  \label{lqcdiso}
\end{equation}
 we have 
\begin{equation}
  h^{0\oplus2}_{\rho^-} =\frac{G_F s_w^2}{3\sqrt{2}}f_\rho m_\rho
     \left(\frac{-C_{5}^{I=0+2}}{3}-C_{6}^{I=0+2}+\frac{C_{7}^{I=0+2}}{3}+C_{8}^{I=0+2}+C_{9}^{I=0+2}+\frac{C_{10}^{I=0+2}}{3}\right)g_A^{u-d}\,.
\end{equation}
With Eqs.(\ref{lqcdiso}), this 
implies
\begin{equation}
         h^{0+2}_{\rho^-} = -18.10 \pm 1.1  + \left(\stackrel{{1.2}}{{}_{-2.4}}\right) + 0.72 + (-4.63))\times 10^{-7} \,.
\end{equation}
Solving Eq.~(\ref{simul}) we find 
\begin{equation}
    h_{\rho}^{0} = - 11.05\pm 0.672  + \left(\stackrel{{1.079}}{{}_{-2.051}}\right) + 0.673 + (-4.039))\times 10^{-7}\,
    ;\\ \quad  h_{\rho}^{2} = + 8.57\pm 0.519  + \left(\stackrel{{1.129}}{{}_{-1.736}}\right) + 0.802 + (-3.749))\times 10^{-7}\,.
\end{equation}
Although our determinations have been made at a scale
of $2\,\rm GeV$, we follow the spirit of DDH~\cite{Desplanques:1979hn} and compare our results
with the constraints on the coupling constants 
that emerge from experiments at
much lower energies. In this way we hope to discern 
the driving 
theoretical limitations 
in our approach. 

\section{Perspectives from comparisons with experiment}
\label{sec:V}
In what follows we consider how the results of 
Sec.~\ref{sec:IV} compare with the outcomes of 
hadronic parity violation experiments with nucleons and nuclei. 
We anticipate that our results may be most closely suited 
to studies of hadronic parity violation in few-body systems, though 
we also consider more complex nuclear systems, 
comparing, in particular, our $h_\pi^1$ result to 
a precise limit extracted from a 
search for parity violation in the radiative 
decay of excited-state $^{18}$F~\cite{Page:1987ak}.
Finally, following earlier work~\cite{Haxton:2013aca,Gardner2017paradigm}, 
we use the DDH potential~\cite{Desplanques:1979hn}, based on one-meson
exchange, 
to evaluate the Danilov parameters
and compare them with the outcomes of low-energy 
experiments, particularly those from parity-violating
proton-proton scattering. We regard these computations
as rough estimates, to be checked against 
the predictions of a large $N_c$ analysis and that 
may serve as guidance in determining the limitations of the DDH potential.

\begin{figure}[!h]
    \centering
    \includegraphics[scale=0.70]{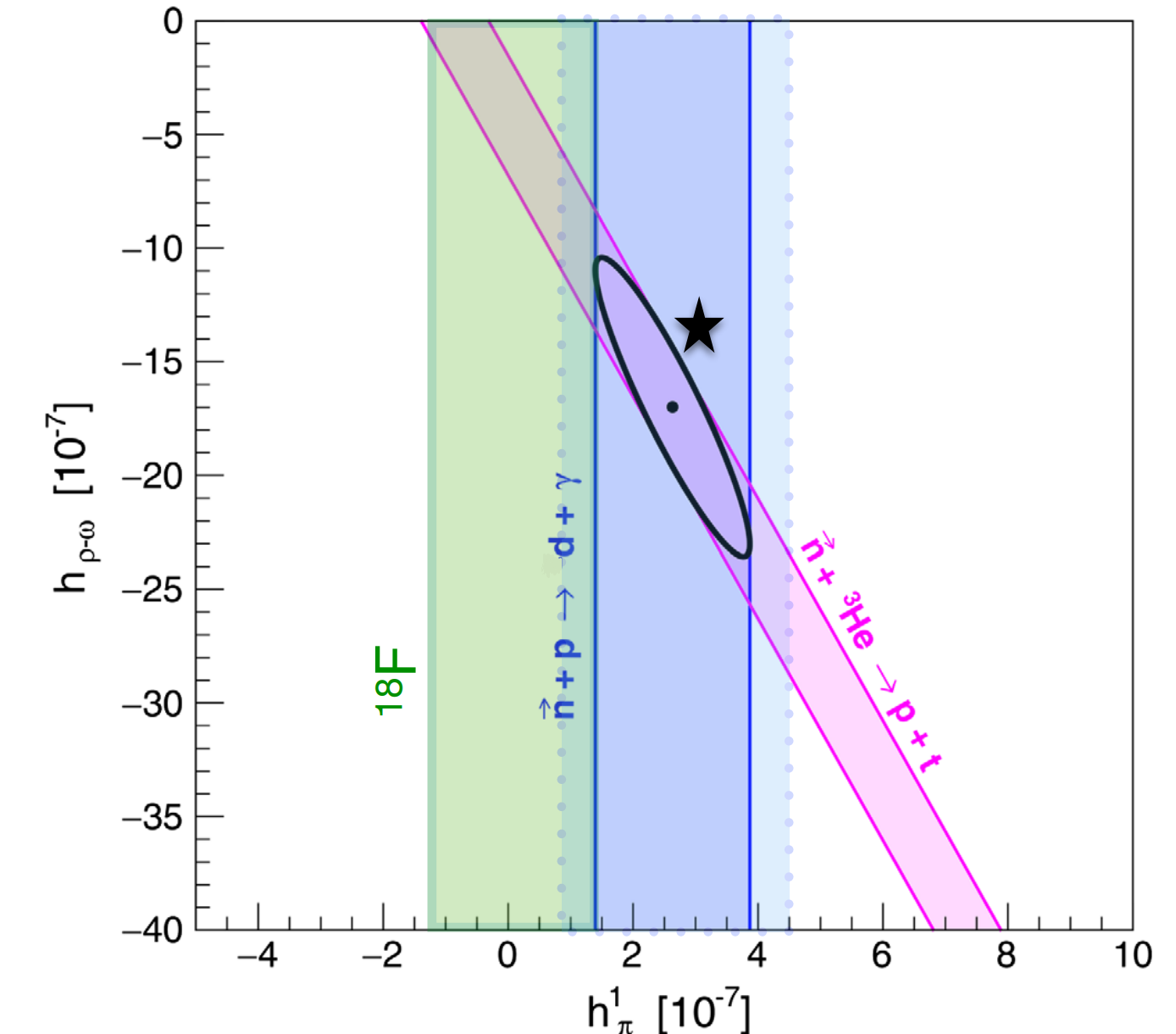}
    \caption{Constraints on the parity-violating 
    coupling constants $h_{\rho-\omega}$ and $h_{\pi}^1$, 
    after 
    Ref.~\cite{n3He:2020zwd}. The couplings are not direct physical 
    observables and thus can be sensitive to the energy scale of the 
    system under consideration, see the text for further discussion. 
    Combining statistical and systematic
    errors in quadrature and working at 68\% CL, 
    we show the value $h_{\pi}^1=(2.6 \pm 1.2) \times 10^{-7}$ 
    from the measured parity-violating asymmetry in ${\vec n} + p\to d + \gamma$~\cite{NPDGamma:2018vhh}
    as the vertical band bounded by a solid line, 
    and its determination $h_{\pi}^1=(2.7 \pm 1.8) \times 10^{-7}$ in chiral perturbation theory 
    as the vertical band bounded by a dotted line~\cite{deVries:2015pza,deVries:2020iea}, and the 
    diagonal constraint from the measured parity-violating asymmetry in $\vec{n} + ^{3}{\rm He}\to p + t$~\cite{n3He:2020zwd}, with the combined fit of the 
    two experiments yielding the ellipse shown. 
The analysis of $^{18}$F radiative decay from its 1.081 MeV excited state
yields the bound $|h_\pi^1| < 1.3 \times 10^{-7}$~\cite{Haxton:2013aca},
shown as the leftmost vertical band. Our \textit{ab initio} result at 
a scale of 
2 GeV is represented by the star with the associated error from its inputs 
roughly by its size. The tension with the $^{18}$F result at a nominal
scale of less than 100 MeV, may also be reflective of an extraction
in a different physical setting. 
}
    \label{fig:hpicf}
\end{figure}

Comparing with the constraints on the 
parity-violating vector-meson-nucleon coupling constants that emerge from the 
combined analysis of the ${\vec n} p \to d \gamma$~\cite{NPDGamma:2018vhh} 
and ${\vec n}\, ^3{\rm He} \to  p\, ^3{\rm H}$~\cite{n3He:2020zwd} experiments, within the theoretical framework of  Ref.~\cite{Viviani:2010qt}, we have 
$h_\pi^1 = (2.6 \pm 1.2_{\rm stat} \pm 0.2_{\rm sys})\times 10^{-7}$~\cite{NPDGamma:2018vhh}, 
and 
$h_{\rho-\omega} \equiv h_\rho^0 + 0.605 h_\omega^0 - 
0.605 h_\rho^1 -1.316 h_\omega^1 + 0.026 h_\rho^2
=(-17.0 \pm 6.56)\times 10^{-7}$~\cite{n3He:2020zwd}, 
for which we compute
\begin{equation}
h_{\rho-\omega} = -12.9 \pm 0.52  + \left(\stackrel{{0.97}}{{}_{-1.9}}\right) + 0.62 + (-3.4))\times 10^{-7} , 
\end{equation}
so that both this and 
our 
$h_\pi^1$, Eq.~(\ref{eq:hpi1}), are within $\pm 1\sigma$ of the experimentally determined parameters. We note, moreover, that analyzing the result of the 
${\vec n} p \to d \gamma$~\cite{NPDGamma:2018vhh} experiment within chiral 
perturbation theory yields $h_\pi^1=(2.7\pm 1.8)\times 10^{-7}$~\cite{deVries:2015pza,deVries:2020iea}. 
Using our results, we evaluate the asymmetry 
in ${\vec n}\, ^3{\rm He} \to p\, ^3{\rm H}$
as $-0.69 \times 10^{-8}$ in the framework of Ref.~\cite{Viviani:2010qt} but as $1.6 \times 10^{-8}$
in the framework of Ref.~\cite{Viviani:2014zha},
as per Eqs.(8,9) of Ref.~\cite{n3He:2020zwd}, to compare
with the experimental result 
$(1.55\pm 0.97_{\rm stat} \pm 0.24_{\rm sys})\times 10^{-8}$~\cite{n3He:2020zwd}. 
Evidently the value of 
the asymmetry is 
sensitive to 
a partial cancellation of the various contributions~\cite{n3He:2020zwd}.
The $h_\pi^1$ determination from the ${\vec n} p \to d \gamma$ experiment, 
$h_\pi^1 = (2.6 \pm 1.2_{\rm stat} \pm 0.2_{\rm sys} )\times10^{-7}$~\cite{NPDGamma:2018vhh} 
is in slight tension with the value determined by the 
non-observation of the 
photon circular polarization in $^{18}$F radiative decay from the 
1.081 MeV $J^P T =0^- 0$ state, reflecting an absence of mixing with the 
nearby 1.042 MeV $0^+ 1$ state, yielding the bound
$|h_\pi^1| < 1.3 \times 10^{-7}$ at 68\% CL~\cite{Haxton:2013aca}. 
The $^{18}{\rm F}$ system is special in that the theoretical uncertainties can be largely 
controlled through the experimental assessment 
of the pertinent nuclear matrix element, after
an isospin rotation, from a 
well-measured
$\beta^+$-decay transition in $^{18}{\rm Ne}$~\cite{Haxton:1981sf,Adelberger:1983zz,Adelberger:1985ik}. 
Thus the error in each $h_\pi^1$ assessment is thought to be statistics dominated. 
Other reliably calculated, parity-violating 
observables that depend on the couplings probed
in the few-body reactions 
include the longitudinal asymmetry in 
elastic $\vec{p}-\alpha$ scattering at $46 \,{\rm MeV}$, 
$A_L [\vec{p}\alpha]$, 
and the gamma asymmetry in $^{19}{\rm F}$ decay, $A_\gamma [^{19}{\rm F}]$. 
Using the expressions in Ref.~\cite{Haxton:2013aca} 
we find 
$-2.6 \times 10^{-7}$, to compare with 
$A_L [\vec{p}\alpha]_{\rm expt}=-(3.3 \pm 0.9)\times 10^{-7}$\cite{Lang,Henneck},  and 
$-6.7 \times 10^{-5}$, to compare with 
$A_\gamma [^{19}{\rm F}]_{\rm expt}=-(7.4 \pm 1.9)\times 10^{-5}$\cite{Adelberger:1983zz,Elsener}.
Therefore 
only the $^{18}$F study is precise enough to challenge the determination of 
$h_\pi^1$ in few-body systems, and we show these results in 
Fig.~\ref{fig:hpicf}, along with the value of $h_\pi^1$ determined from the 
parity-violating gamma asymmetry in ${\vec n} p \to d \gamma$~\cite{NPDGamma:2018vhh}
using chiral perturbation theory~\cite{deVries:2015pza,deVries:2020iea}, as well as our own 
determination of that and of $h_{\rho-\omega}$. Our assessment
of these couplings at a renormalization scale of $\mu=2$ GeV is compatible
with the determinations from the few-body results, but both it and
the experiment values are in tension with the
$^{18}$F result. Of course it is possible that the disagreement 
between the experiments could be 
experimental in origin, though the procedures used in the NPDGamma experiment
have been validated through the experimental study of parity-violating 
$\vec{n}$ capture on 
$^{35}$Cl~\cite{Fomin:2022eqn}, or be the result of an underestimated
theoretical systematic error, yet 
we emphasize that these couplings are not directly observable. 
Thus they 
can be expected to vary with the 
renormalization scale of the
system in which they are determined, which is typically bounded from above by 
the cutoff scale
that determines the active degrees of freedom in a particular EFT. 
In the current context we contrast chiral perturbation theory, a NN EFT with active pion degrees of
freedom and a cutoff scale of about 1 GeV~\cite{Gasser:1984gg,Bernard:2006gx}, with chiral effective theory, 
an EFT in which pion degrees of freedom are absent and thus with 
a cutoff scale of about 100 MeV. 
In settings where the scale variation is set by perturbative
physics, such as in the case of the running of $\sin^2 \theta_W$ 
in the SM, noting Fig.~5 of Ref.~\cite{Carlini:2019ksi}, in which 
the natural scale choice 
is the typical momentum transfer $Q$ of the experiment, 
the computed variations 
are numerically very small, a few percent at most. 
However, in low-energy QCD, the scale variation 
is no longer controlled by weakly-coupled effects, 
and it need not be very small. 
To illustrate, 
we turn to a NN 
effective theory 
without pions, 
so-called 
pionless effective 
theory~\cite{Chen:1999tn,vanKolck:1999mw,Beane:2000fx}.
The large $S$-wave scattering lengths $a_0^J$, with $J=0,1$, associated with the low-energy NN system reflect
the possibility of nearly or weakly bound states, and to address 
the incompatibility of that large length scale in an effective theory with 
a break-down scale of $\Lambda_{\slash{\!\!\!\pi}}$\cite{Kaplan:1996xu,Kaplan:1998tg,Kaplan:1998we,vanKolck:1998bw}, where $a_0^J \gg 1/\Lambda_{\slash{\!\!\!\pi}}$, a power-divergence 
subtraction (PDS) scheme 
can be employed at a subtraction point of $\mu \approx Q$~\cite{Kaplan:1998tg}.
In this scheme the 
LECs that result vary 
with $\mu$ as a ratio of simple polynomials, and we note 
that the $\mu$ variation in ratios of LECs can vary by a factor of a few over scales $\mu$ ranging from 80 to 180 MeV~\cite{Schindler:2018irz}. 
Although the PDS
scheme enlarges the range of momenta for which the EFT is valid, 
other, long-standing approaches to the systematic organization of 
a chiral EFT continue to be followed~\cite{Weinberg:1990rz}.
We note Ref.~\cite{Epelbaum:2017byx} for a detailed comparative
study of the 
PDS renormalization and 
the Wilsonian renormalization group schemes in an analytically solvable NN EFT; here we
consider the implications of their conjecture that fitting 
LECs 
to a data set implicitly selects a renormalization scheme. 
To us, this means the particular parity-violating 
couplings shown in Fig. ~\ref{fig:hpicf} can intrinsically depend on the
physical momentum scale of the studies in which they are extracted. 
Here we note that a cutoff scale of the EFT that would describe 
the radiative decay of an excited state of $^{18}$F, 
which is pertinent even if the existing extraction is regarded
as semi-empirical~\cite{Adelberger:1983zz,Adelberger:1985ik,Haxton:1981sf}, 
is much lower than the one associated with 
chiral perturbation theory for ${\vec n} + p \to d + \gamma$. 
The extracted couplings could be discernibly different 
in the two settings, and we consider probes of this possibility
in what follows.

Recent analyses 
have suggested that matrix elements of a quark-based 
effective Hamiltonian can be matched to chiral perturbation
theory at a renormalization scale of $\mu=2\,\rm GeV$~\cite{Cirigliano:2018yza,Cirigliano:2022rmf}. 
Conventionally, however, the cutoff scale of chiral 
perturbation theory is taken to be $1\,\rm GeV$~\cite{Gasser:1983yg,Bernard:2006gx}, or
the $\rho$ mass~\cite{Aoki:2021kgd}. 
If we were to try to evolve our description to still lower scales, 
we expect to encounter the charm quark 
scale at $\mu=m_c$~\cite{Marciano:1983pj}.  
For $\mu \gg m_c \approx 1.3\,\rm GeV$, the effects of the charm-quark mass
are negligible, allowing $u$-like quark penguin contributions from 
the charged-current contributions in the weak effective Hamiltonian 
to cancel. However, at scales 
for which $\mu\gtrsim m_c$, this cancellation is no longer 
efficient, and if $\mu \le m_c$, it no longer operates. 
Thus for $\mu < 2\,\rm GeV$ the effects of these 
additional operators, all of $I=0$ character, can exist~\cite{Gardner:2022mxf}, 
along with 
the possibility of non-perturbative matching~\cite{Tomii:2020smd}
that we have already noted. These effects are presumably small with respect
to the precision of the $h_\pi^1$ extraction from chiral perturbation
theory~\cite{deVries:2015pza,deVries:2020iea}, nominally at a scale
of $\mu=1\,\rm GeV$, shown in Fig.~\ref{fig:hpicf}. 
Nevertheless, to begin to assess the possible numerical implications of these effects, 
we use the coupling constants we have computed as they stand 
to estimate the LECs of very-low-energy, parity-violating observables
in the NN system, which are essentially the Danilov parameters~\cite{Zhu:2004vw}, 
to compare more broadly with existing experiments. 
Working within the context of the DDH potential, with parameters
$g_{\pi NN}^2/4\pi =14.4$,
$g_{\rho}^2/4\pi =0.62$, $g_{\omega}^2/4\pi =9g_{\rho}^2/4\pi$,
$\chi_\rho=3.70$, and $\chi_\omega=-0.12$, we
compute the Danilov parameters 
to find
\begin{equation}
\begin{split}
&\Lambda_0^{^1S_0-^3P_0}=-g_\rho(2+\chi_\rho)h_\rho^0-g_\omega(2+\chi_\omega)h_\omega^0 
\to 176 \,[210]   \\
&\Lambda_0^{^3S_1-^1P_1}=-3g_\rho\chi_\rho h_\rho^0+g_\omega\chi_\omega h_\omega^0
 \to 343\, [360]
\\
&\Lambda_1^{^1S_0-^3P_0}=-g_\rho(2+\chi_\rho  )h_\rho^1-g_\omega(2+\chi_\omega )h_\omega^1
 \to 4.67 \, [21] 
\\
&\Lambda_1^{^3S_1-^3P_1}=\frac{g_{\pi N N}}{\sqrt{2}}\left({\frac{m_\rho}{m_\pi}}\right)^2\!h_\pi^1
+g_\rho(h_\rho^1 -{h_\rho^1}')
-g_\omega h_\omega^1
 \to 859 \, [1340]  
 \\
&\Lambda_2^{^1S_0-^3P_0}= -g_\rho(2+\chi_\rho )h_\rho^2
 \to -137\, [160] \,, 
\end{split}
 \label{eq:danilov}
\end{equation}
where we neglect ${h_\rho^1}'$~\cite{Holstein:1981, Haxton:2013aca} and provide
our numerical values, with 
the DDH ``best values~\cite{Desplanques:1979hn}'' given in brackets --- and all in units of $10^{-7}$.
Following the large $N_c$ analysis of Ref.~\cite{Gardner2017paradigm}, we compute 
\begin{equation}
\Lambda_0^+ \equiv \frac{1}{4}\Lambda_0^{^1S_0-^3P_0}
+ \frac{3}{4}\Lambda_0^{^3S_1-^1P_1} \to  301 \,\,;\,\,
\Lambda_0^- \equiv  \frac{1}{4}\Lambda_0^{^3S_1-^1P_1}
-  \frac{3}{4}\Lambda_0^{^1S_0-^3P_0} \to  -46 \,, 
\end{equation}
and recall the scaling predictions 
$\Lambda_0^+\sim N_c$, $\Lambda_2^{^1S_0-^3P_0}  \sim N_c\sin^2 \theta_w$,
$\Lambda_0^-\sim 1/N_c$, $\Lambda_1^{^1S_0-^3P_0}  \sim \sin^2 \theta_w$, 
$\Lambda_1^{^3S_1-^3P_1}  \sim \sin^2 \theta_w$~\cite{Phillips:2014kna,Schindler:2015nga,Gardner2017paradigm}. Certainly the value of $h_\pi^1$ we
compute yields a value of $\Lambda_1^{^3S_1-^3P_1}$ at odds with the large $N_c$ expectation, 
though $\Lambda_1^{^3S_1-^3P_1}|_{h_\pi^1=0}= -31$. 
We note, too, that in this we have ignored
the possibility of scale dependence entirely, though 
an explicit study~\cite{Schindler:2018irz} 
in the parity-conserving case shows that 
only certain ranges of $\mu$ are compatible 
with large $N_c$ expectations for partial waves beyond the $S$-wave channels. 

We now turn to other observables, starting with the 
parity-violating longitudinal asymmetry in low-energy 
$\vec{p}p$ scattering, $A_L (\vec{p} p)$, for which the Danilov parameters associated 
with $S-P$ interference should suffice. Fixed target 
$\vec{p}p$ experiments 
at beam energies of 13.6 MeV, 15 MeV, and 45 MeV can be 
analyzed within a DDH framework~\cite{Carlson:2001ma} to 
yield~\cite{Haxton:2013aca}
\begin{equation}
\frac{2}{5} \Lambda_0^+ + \frac{1}{\sqrt{6}} \Lambda_2^{^1S_0-^3P_0} + \left[ 
 \Lambda_1^{^1S_0-^3P_0} -\frac{6}{5} \Lambda_0^- \right] 
= 419\pm 43 \,,
\end{equation}
which we evaluate as $120 - 56 + 60 =124$.
Thus our results in this case do not compare favorably. 
For context, we note that 
an analysis of this observable in chiral 
effective theory shows that correlated 
two-pion exchange (TPE) also 
plays an important role~\cite{deVries:2013fxa,Viviani:2014zha,deVries:2014vqa}, 
bringing in an interaction largely controlled
by $h_\pi^1$ as well, although TPE is not present in  
the DDH framework. 
As for the other observables we have
considered, the value of $h_\pi^1$ plays an 
important numerical role, with the subleading
contributions, which are largely isovector, and
the leading ones, which are isoscalar, playing
comparable numerical roles. Thus 
although our original assessment of the Danilov 
parameters, with the exception of the one 
in which $h_\pi^1$ appears,
are crudely consistent with large $N_c$ scaling, 
it appears that 
the large $N_c$ 
relationships are not effective in predicting the aggregate
size of the various contributions. In this the 
parameter $h_\pi^1$ drives this conclusion, making
its computation within LQCD~\cite{Feng:2017iqb,Sen:2021dcb}, 
noting the pioneering work of Ref.~\cite{Wasem:2011tp}, or an 
improved experimental assessment of it, possibly through a
next-generation 
$\vec{n} p \to d \gamma$ experiment, extremely 
welcome. 
Another interesting possibility would
be a neutron spin 
rotation experiment in 
liquid $^4$He; the existing 
limit is consistent with zero but is statistics limited~\cite{Swanson:2019cld}, and 
 a new experiment with a planned factor of 10 improvement in
sensitivity is being developed~\cite{NIST2019}. With our Danilov parameter estimates that experiment should be able to measure a non-zero result.
As for our suggestion that the 
extraction of $h_\pi^1$, and possibly other couplings, 
could vary with the cutoff scale of the physical description, 
we hope that further studies of hadronic parity 
violation in complex systems could be made and be of sufficient
precision to reveal this effect in other isosectors as well. 
Since we have noted that additional penguin contributions, of purely 
isoscalar character, emerge once the charm quark is no
longer an active degree of freedom, we think that precision 
experimental studies of hadronic parity violation 
in the isoscalar sector, as detailed in Ref.~\cite{Gardner2017paradigm}, 
both in few-body and complex nuclei, would be needed to 
assess the quantitative 
importance of these long-neglected effects. 
A particularly appealing example would be the measurement of 
the parity-violating asymmetry in 
$\vec{n} + d \to t + \gamma$, because the asymmetry is expected to be 
somewhat larger than those of other measured reactions, with little 
sensitivity to the isotensor sector 
--- and it would be interesting to compare that outcome to the 
measured
$\gamma$-ray asymmetry in $^{19}$F decay~\cite{Haxton:2013aca,Gardner2017paradigm}
and even more so if the precision of the latter experiment could be improved. 

\section{Summary} 
\label{sec:VI}
We have used the LO QCD effective weak Hamiltonian for parity-violating, $\Delta S = 0$ hadronic processes
to determine the parity-violating meson-nucleon coupling constants, $h_\pi^1$, $h_\rho^{0,1,2}$, 
$h_\omega^{0,1}$, familiar from the DDH framework. We have achieved this by
employing the 
factorization {\it Ansatz} and 
assessments of the pertinent quark charges of the nucleon in lattice QCD at the $2\,\rm GeV$ scale. Working further, we have found that our assessment of $h_\pi^1$ and $h_{\rho-\omega}$ agree within $1\sigma$ of their experimental determinations in few-body
nuclear systems~\cite{NPDGamma:2018vhh,n3He:2020zwd},
though both our $h_\pi^1$ result and the size of the 
asymmetry in $\vec{n} p \to d \gamma$~\cite{NPDGamma:2018vhh}
are in slight tension with the null result from the study of 
$P_\gamma [^{18}{\rm F}]$~\cite{Haxton:1981sf,Adelberger:1983zz,Adelberger:1985ik}, and we have noted the possibility that the extracted coupling could depend on the cutoff scale of the EFT description that would describe it. 

 Turning to the study of the parity-violating asymmetries 
in low-energy $\vec{p}p$ scattering, which is 
sensitive to the $I=2$ Davilov parameter $\Lambda_2^{^1S_0-^3P_0}$ as well, 
we do not find agreement with experiment. The analysis
of this process within chiral effective theory, however, 
suggests that TPE, an effect not included in the DDH potential, plays an important role~\cite{Viviani:2014zha},
 and this can also modify the $I=1$ Danilov parameters,
 though it may be that our factorization 
 assessment of $h_\rho^2$, or of neglected higher order
 effects in $\alpha_s$, and thus of 
 $\Lambda_2^{^1S_0-^3P_0}$ that is to blame. 
 We note that the 
 parameter $h_{\rho-\omega}$ depends only very weakly on 
the $I=2$ sector. 

Five independent parameters characterize low-energy
hadronic parity violation, and the use of 
pionless effective theory in the large $N_c$ limit gives 
insight into the relative size of the 
contributions~\cite{Zhu:2009,Phillips:2014kna,Schindler:2015nga,Gardner2017paradigm}. Yet these are scaling relationships, 
rather than numerical predictions, and we have noted
that our numerical assessments in Eq.~(\ref{eq:danilov}), 
save for the $I=1$ parameter containing $h_\pi^1$, 
compare favorably with those expectations. 
Thus the overall success of the large $N_c$ predictions 
very much depends on the precise value of $h_\pi^1$, with future 
input from either LQCD or experiment important to a definitive test. 
Despite this, the application of 
our results, within the DDH framework, to 
parity-violating observables in $A>3$ systems
suggest that it is not effective, because the 
subleading pieces are not only quite large, but they are
also needed for theoretical compatibility with the 
observed effects. 
This outcome is nevertheless
suggestive that
the systematic study of 
hadronic parity violation in $A>3$ systems, 
for which studies in molecular 
systems~\cite{Borschevsky:2012wp} also show great 
promise~\cite{Altuntas:2018ots,Karthein22}, is within reach. 
Precision experimental studies, particularly in the isoscalar sector, 
can illuminate the additional theoretical 
effects we have noted, 
providing an important opportunity 
to bench-mark end-to-end EFT descriptions of low-energy 
weak observables in nuclei, which play a broad role 
in searches for physics beyond the SM.

The hints of success in our work of updating DDH come from 
comparing 
our 
estimations of meson-nucleon couplings
with the outcomes of recent experiments, as discussed at length in the current paper. We are careful not to claim any such comparisons 
with our crude estimations of 
the Danilov parameters. 
Yet, we are of the opinion that such rough estimations can be checked against the predictions of 
large $N_c$ analysis and may serve as a supplement in assessing the limitations 
of our approach. 
We would like to emphasize that our work neither discounts the possibility 
of TPE
nor of the importance of additional 
non-perturbative effects in a complete picture of hadronic parity 
violation at low energies. But, in striving 
to refine 
the benchmark expectations of the
parity-violating meson-nucleon 
couplings, 
we have updated the work of
DDH~\cite{Desplanques:1979hn}
via the introduction of renormalization-group 
methods, a modern definition of factorization, and lattice QCD inputs 
and thus in so doing 
overcome many challenges in such theoretical computations starting 
in the 
1980s.
Future theoretical work that would 
aspire to confront low-energy 
experiments more directly would surely benefit from the 
realization of a LQCD program for the computation of 
2N matrix elements for hadronic parity violation, which is under development~\cite{Kurth:2015cvl,Horz:2020zvv,Walker-Loud22}, though there
are ongoing challenges~\cite{Nicholson:2021zwi}.

\section*{Acknowledgments}
We acknowledge partial support from the U.S. Department of Energy Office
of Nuclear Physics under contract DE-FG02-96ER40989.
We thank the INT for gracious hospitality and 
the workshop participants 
of ``Hadronic Parity Nonconservation II'' for helpful
discussions during the early stages of this work. 

\appendix

\section{Four-quark Operators}
\label{appendix A}
The operators of the complete theory ($\mathcal{H}_{\rm eff}^{\rm PV}$) with all three isosectors are:
\begin{equation}\label{full ops}
\begin{split}
   \Theta_1 & = [(\bar{u}u)_V+(\bar{d}d)_V+(\bar{s}s)_V]^{\alpha\alpha}[(\bar{u}u)_A-(\bar{d}d)_A-(\bar{s}s)_A]^{\beta\beta}\\
   \Theta_2 & = [(\bar{u}u)_V+(\bar{d}d)_V+(\bar{s}s)_V]^{\alpha\beta}[(\bar{u}u)_A-(\bar{d}d)_A-(\bar{s}s)_A]^{\beta\alpha}\\
   \Theta_3 & = [(\bar{u}u)_A+(\bar{d}d)_A+(\bar{s}s)_A]^{\alpha\alpha}[(\bar{u}u)_V-(\bar{d}d)_V-(\bar{s}s)_V]^{\beta\beta}\\
   \Theta_4 & = [(\bar{u}u)_A+(\bar{d}d)_A+(\bar{s}s)_A]^{\alpha\beta}[(\bar{u}u)_V-(\bar{d}d)_V-(\bar{s}s)_V]^{\beta\alpha}\\
  \Theta_5 & = [(\bar{u}u)_V-(\bar{d}d)_V-(\bar{s}s)_V]^{\alpha\alpha}[(\bar{u}u)_A-(\bar{d}d)_A-(\bar{s}s)_A]^{\beta\beta}\\
  \Theta_6 & = [(\bar{u}u)_V-(\bar{d}d)_V-(\bar{s}s)_V]^{\alpha\beta}[(\bar{u}u)_A-(\bar{d}d)_A-(\bar{s}s)_A]^{\beta\alpha}\\
  \Theta_7 & = [(\bar{u}u)_A+(\bar{d}d)_A+(\bar{s}s)_A]^{\alpha\alpha}[(\bar{u}u)_V+(\bar{d}d)_V+(\bar{s}s)_V]^{\beta\beta}\\
  \Theta_8 & = [(\bar{u}u)_A+(\bar{d}d)_A+(\bar{s}s)_A]^{\alpha\beta}[(\bar{u}u)_V+(\bar{d}d)_V+(\bar{s}s)_V]^{\beta\alpha} \\
    \Theta_9 &= (\bar{u}d)_V^{\alpha \alpha}(\bar{d}u)_A^{\beta \beta}+(\bar{d}u)_V^{\alpha \alpha}(\bar{u}d)_A^{\beta \beta}\\
   \Theta_{10} &= (\bar{u}d)_V^{\alpha \beta}(\bar{d}u)_A^{\beta \alpha}+(\bar{d}u)_V^{\alpha \beta}(\bar{u}d)_A^{\beta \alpha}\\
    \Theta_{11} &= (\bar{u}s)_V^{\alpha \alpha}(\bar{s}u)_A^{\beta \beta}+(\bar{s}u)_V^{\alpha \alpha}(\bar{u}s)_A^{\beta \beta}\\
    \Theta_{12} &= (\bar{u}s)_V^{\alpha \beta}(\bar{s}u)_A^{\beta \alpha}+(\bar{s}u)_V^{\alpha \beta}(\bar{u}s)_A^{\beta \alpha} \,.
\end{split}
\end{equation}

Operators for isovector sector ($\mathcal{H}_{\rm eff}^{\rm I=1}$) are:
\begin{equation}\label{isovec op set}
    \begin{split}
        \Theta_1&^{I=1}= [(\bar{u}u)_V+(\bar{d}d)_V+(\bar{s}s)_V]^{\alpha\alpha}[(\bar{u}u)_A-(\bar{d}d)_A]^{\beta\beta}\\
        \Theta_2&^{I=1}= [(\bar{u}u)_V+(\bar{d}d)_V+(\bar{s}s)_V]^{\alpha\beta}[(\bar{u}u)_A-(\bar{d}d)_A]^{\beta\alpha}\\
        \Theta_3&^{I=1}= [(\bar{u}u)_A+(\bar{d}d)_A+(\bar{s}s)_A]^{\alpha\alpha}[(\bar{u}u)_V-(\bar{d}d)_V]^{\beta\beta}\\
        \Theta_4&^{I=1}= [(\bar{u}u)_A+(\bar{d}d)_A+(\bar{s}s)_A]^{\alpha\beta}[(\bar{u}u)_V-(\bar{d}d)_V]^{\beta\alpha}\\
        \Theta_5&^{I=1}= (\bar{s}s)_V^{\alpha\alpha}[(\bar{u}u)_A-(\bar{d}d)_A]^{\beta\beta}\\
        \Theta_6&^{I=1}= (\bar{s}s)_V^{\alpha\beta}[(\bar{u}u)_A-(\bar{d}d)_A]^{\beta\alpha}\\
        \Theta_7&^{I=1}= (\bar{s}s)_A^{\alpha\alpha}[(\bar{u}u)_V-(\bar{d}d)_V]^{\beta\beta}\\
        \Theta_8&^{I=1}= (\bar{s}s)_A^{\alpha\beta}[(\bar{u}u)_V-(\bar{d}d)_V]^{\beta\alpha}\\
        \Theta_9&^{I=1}= (\bar{u}s)_V^{\alpha\alpha}(\bar{s}u)_A^{\beta\beta} + (\bar{s}u)_V^{\alpha\alpha}(\bar{u}s)_A^{\beta\beta}\\
        \Theta_{10}&^{I=1}= (\bar{u}s)_V^{\alpha\beta}(\bar{s}u)_A^{\beta\alpha} + (\bar{s}u)_V^{\alpha\beta}(\bar{u}s)_A^{\beta\alpha}\\
        \end{split} \,,
\end{equation}
and the operators for $I=0 \oplus 2$ sector  ($\mathcal{H}_{\rm eff}^{\rm I=0 \oplus 2}$) are:
\begin{equation}\label{isoeven ops}
    \begin{split}
        \Theta_1&^{I=0 \oplus 2}= [(\bar{u}u)_V+(\bar{d}d)_V+(\bar{s}s)_V]^{\alpha\alpha}[(\bar{s}s)_A]^{\beta\beta}\\
        \Theta_2&^{I=0 \oplus 2}= [(\bar{u}u)_V+(\bar{d}d)_V+(\bar{s}s)_V]^{\alpha\beta}[(\bar{s}s)_A]^{\beta\alpha}\\
         \Theta_3&^{I=0 \oplus 2}= [(\bar{u}u)_A+(\bar{d}d)_A+(\bar{s}s)_A]^{\alpha\alpha}[(\bar{s}s)_V]^{\beta\beta}\\
        \Theta_4&^{I=0 \oplus 2}= [(\bar{u}u)_A+(\bar{d}d)_A+(\bar{s}s)_A]^{\alpha\beta}[(\bar{s}s)_V]^{\beta\alpha}\\
        \Theta_5&^{I=0 \oplus 2}= [(\bar{u}u)_V-(\bar{d}d)_V]^{\alpha\alpha}[(\bar{u}u)_A-(\bar{d}d)_A]^{\beta\beta}+(\bar{s}s)_V^{\alpha\alpha}(\bar{s}s)_A^{\beta\beta}\\
        \Theta_6&^{I=0 \oplus 2}= [(\bar{u}u)_V-(\bar{d}d)_V]^{\alpha\beta}[(\bar{u}u)_A-(\bar{d}d)_A]^{\beta\alpha}+(\bar{s}s)_V^{\alpha\beta}(\bar{s}s)_A^{\beta\alpha}\\
       \Theta_7&^{I=0 \oplus 2}= [(\bar{u}u)_V+(\bar{d}d)_V+(\bar{s}s)_V]^{\alpha\alpha}[(\bar{u}u)_A+(\bar{d}d)_A+(\bar{s}s)_A]^{\beta\beta}\\
       \Theta_8&^{I=0 \oplus 2}= [(\bar{u}u)_A+(\bar{d}d)_A+(\bar{s}s)_A]^{\alpha\beta}[(\bar{u}u)_V+(\bar{d}d)_V+(\bar{s}s)_V]^{\beta\alpha}\\
        \Theta_9&^{I=0 \oplus 2}= (\bar{u}d)_V^{\alpha\alpha}(\bar{d}u)_A^{\beta\beta} + (\bar{d}u)_V^{\alpha\alpha}(\bar{u}d)_A^{\beta\beta}\\
        \Theta_{10}&^{I=0 \oplus 2}= (\bar{u}d)_V^{\alpha\beta}(\bar{d}u)_A^{\beta\alpha} + (\bar{d}u)_V^{\alpha\beta}(\bar{u}d)_A^{\beta\alpha}\\
        \end{split} \,.
\end{equation}

\bibliography{main_prc_version3.bbl}

\end{document}